\documentclass[nofootinbib]{revtex4}
\usepackage{graphicx}
\usepackage{latexsym}
\def\be{\begin{equation}}
\def\ee{\end{equation}}
\def\bea{\begin{eqnarray}}
\def\eea{\end{eqnarray}}

\begin{document}

\title{Metric-Affine Formalism of Higher Derivative Scalar Fields in Cosmology}
\author{Mingzhe Li}
\email{limz@nju.edu.cn}
\author{Xiulian Wang}
\email{wangxiulian2000@yahoo.com.cn}

\affiliation{Department of
Physics, Nanjing University, Nanjing 210093, P.R. China}
\affiliation{Joint Center for Particle,
Nuclear Physics and Cosmology, Nanjing University - Purple Mountain
Observatory, Nanjing 210093, P.R. China}


\begin{abstract}

Higher derivative scalar field theories
have received considerable attention for the potentially explanations of the initial state of the universe or the current cosmic acceleration which
they might offer. They have also attracted many interests in the phenomenological studies of infrared modifications of gravity. These theories
are mostly studied by the metric variational approach in which only the metric is the fundamental field to account for the gravitation. In this
paper we study the higher derivative scalar fields with the metric-affine formalism where the connections are treated arbitrarily at the beginning.
Because the higher derivative scalar fields couple to the connections directly in a covariant theory these two formalisms will lead to different results.
These differences are suppressed by the powers of the Planck mass and are usually expected to have small effects. But in
some cases they may cause non-negligible deviations. We show by a higher derivative dark energy model that the two formalisms lead to
significantly different pictures of the future universe.

\end{abstract}

\maketitle

\hskip 1.6cm PACS number(s): 98.80.Cq. \vskip 0.4cm

\section{Introduction}

The scalar fields have extensive applications in cosmology. They were used in building the early universe models,
 such as inflation \cite{Guth:1980zm,Linde:1981mu,Albrecht:1982wi},
bouncing cosmology \cite{Novello:2008ra}, cyclic universe \cite{Steinhardt:2002ih}, pre-big-bang \cite{Gasperini:1992em} and so on to
generate the initial conditions of the structure formation.
They were also modeled as
the mysterious dark energy to drive the current accelerating expansion of the universe \cite{Copeland:2006wr}.
Among these applications, the higher derivative scalar field models attracted special interests.
The non-degenerate higher derivative theories (in which the Lagrangian density depends on the higher derivative terms non-linearly)
were proposed in Ref. \cite{Li:2005fm} to construct single field quintom
dark energy models \cite{Feng:2004ad} which cross the cosmological constant boundary during evolution and violate the null energy condition.
Such theories were also applied in the early universe to build inflation models \cite{Anisimov:2005ne}, and, due to the property of
violating the null energy condition, to realize the non-singular bouncing cosmology \cite{Cai:2007qw,Cai:2008qw}. Such non-degenerate
higher derivative scalar field models are plagued by the presence of the ghost degrees of freedom which lead to the quantum instability
even though they can be classically stable.
Another motivation to study the higher derivative scalar field theories is stimulated by the recently proposed
Galileon models \cite{Nicolis:2008in}. The Galileon models are viewed as local infrared
modifications of general relativity (GR), generalizing an effective field
description of the DGP model \cite{Dvali:2000hr}. Specifically, the Galileons are higher derivative scalar fields that
nevertheless have second-order equations of motion
and are invariant under the ``galilean" symmetry, $\phi(x)\rightarrow \phi(x)+c+b_{\mu}x^{\mu}$, where $c$ is a constant and $b_{\mu}$ are
the components of a constant vector.
The second order equations of motion for the Galileons make the theory free from propagating extra ghostly degrees of freedom.
Strictly speaking, this picture is only valid in the Minkowski spacetime. In the curved spacetime, not all the
higher derivative terms in originally proposed Lagrangian density for
the Galileon \cite{Nicolis:2008in} can guarantee a second order equation of motion.
To eliminate the higher order derivatives in the equation of motion, some non-minimal couplings between the curvature and
the scalar field have to be introduced \cite{Deffayet:2009wt}. However, in this resulting theory that
was in fact discovered by Horndeski about four decades ago \cite{Horndeski:1974}, the galilean symmetry is broken.
After relaxing the galilean symmetry and without considering the non-minimal couplings to the curvature, the higher derivative scalar fields
which only have second order equations of motion must have degenerate Lagrangian densities, that means the Lagrangian densities depend on the
higher derivative terms linearly \cite{Li:2011qf}. Such type of models present novel interesting phenomena. They correspond to
imperfect fluids \cite{Deffayet:2010qz,Pujolas:2011he} and more importantly may violate the null energy condition without any instability
in the fluctuations.
Due to this advantage, they are used to realize the quintom dark energy scenario \cite{Li:2011qf,Deffayet:2010qz} and to
address issues in the early universe through inflation \cite{Kobayashi:2010cm} and non-singular bounce \cite{Qiu:2011cy}.
Some other phenomenological studies of the Galileon or generalized Galileon models can be found in Refs. \cite{deRham:2010eu, Deffayet:2009mn,
Gao:2011qe,Chow:2009fm, Silva:2009km, DeFelice:2010pv,
Hinterbichler:2010xn, Creminelli:2010ba}.

Usually these higher derivative scalar field theories are studied through the metric approach.
Like in standard GR, the spacetime is assumed to be pseudo-Riemannian and the metric is the only fundamental field to account for the gravitation.
The connections have a prior relationship with
the metric tensor, $\Gamma^{\mu}_{~\rho\sigma}=\frac{1}{2}g^{\mu\nu}(\partial_{\rho}g_{\nu\sigma}+\partial_{\sigma}g_{\rho\nu}
-\partial_{\nu}g_{\rho\sigma})$ which are called as the Levi-Civita connections or the Christoffel symbols in the literature. All the variables
like the curvature and the covariant derivative operators are defined by these Levi-Civita connections and the metric tensor. To derive the
gravitational field equations from the variational principle, we need only vary the given action with respect to the metric.

As we know, there is an alternative variation approach in which the connections and the metric are taken as independent quantities.
To get the field equations, one must vary the action with respect to both of them. This approach is called as the Palatini formalism or more
precisely the  metric-affine formalism and studied many times in the literature,
see for example \cite{Hehl1,Hehl2,Vollick:2003aw,Sotiriou:2006qn}. In the differential geometry, the connections are used to define the parallel
transports of the tensor fields in the manifold and have
independent geometric meanings. So the metric-affine formalism is usually thought to be more natural than the metric approach.
If the matter fields minimally couple to the metric and do not couple to the connections directly, and the action of gravity is
the Einstein-Hilbert type, the metric-affine formalism lead to the same equations with GR. However, in the following cases,
two approaches will predict different dynamics: (i) modified gravity theories, for example the extensively studied $f(R)$ theories
\cite{Vollick:2003aw,Sotiriou:2006qn};
(ii) non-minimal couplings between the matter fields and the gravitational field \cite{Bauer:2008zj};
(iii) the matter fields couple to the connections directly,
for example when the spinor fields are present, their Lagrangian densities should contain the connections explicitly \cite{weyl}.

The higher derivative scalar fields must also couple to the connections directly as a result of covariant theories.
So using the metric-affine formalism, some differences from the results studied through the metric approach are expected.
These differences may be taken as to account for the deviations of the spacetime from pseudo-Riemannian or as modifications to the scalar
field equation. We will take the second viewpoint and derive the effective Lagrangian density of the scalar field from the metric-affine formalism,
from which we can see that the modification terms are suppressed by the powers of the Planck mass and usually have small effects.
However, in some cases where
the original Lagrangian density itself contains the higher order terms these modification terms could be significant.

This paper is organized as follows. In Section II, we simply review the metric-affine formalism; in Section III we study in detail what would be
the differences produced by the metric-affine formalism in the higher derivative scalar field theories; in Section IV we illustrate the significance
of the modifications brought by the metric-affine formalism in a simple model; Section V is the summary.

\section{The metric-affine formalism}

As mentioned in the Introduction, with the metric variational principle, the spacetime is assumed to be pseudo-Riemannian throughout
and fully specified by the symmetric metric tensor. The affine connections are chosen to be a priori the Levi-Civita connections,
which are symmetric $\Gamma^{\mu}_{~\rho\sigma}=\Gamma^{\mu}_{~\sigma\rho}$ and metric compatible $\nabla_{\mu}g_{\rho\sigma}=0$.
The gravitational field equation is obtained by varying the action only with respect to the metric. In the standard GR,
we adopt the Einstein-Hilbert action for the gravity,
\be\label{e}
S_g=\frac{M_P^2}{2}\int d^4 x\sqrt{g} R~,
\ee
where $M_P^2\equiv 1/\sqrt{8\pi G}$ is the reduced
Planck mass and $g=-{\rm det}|g_{\mu\nu}|$, $R$ is the
curvature scalar which is finally written as a function of the metric and its derivatives via the relations of the Levi-Civita
connections to the metric.

However, in the metric-affine approach, the connections are treated independently, they have no a priori relationship with the metric tensor.
We will use symbols with hats to distinguish the variables and operators associated with the general connections from those with the Levi-Civita
connections. First we use these general connections to define the covariant derivative
\be\label{covariant}
\hat{\nabla}_{\rho}T^{\mu}_{\nu}=\partial_{\rho}T^{\mu}_{\nu}+\hat{\Gamma}^{\mu}_{\rho\sigma}T^{\sigma}_{\nu}
-\hat{\Gamma}^{\sigma}_{\rho\nu}T^{\mu}_{\sigma}~,
\ee
where $T^{\mu}_{\nu}$ is an arbitrary tensor field. The general connections are not limited to be symmetric, the antisymmetric part is called
as the torsion tensor
\be
S^{\mu}_{~\rho\sigma}\equiv \frac{1}{2}(\hat{\Gamma}^{\mu}_{~\rho\sigma}-\hat{\Gamma}^{\mu}_{~\sigma\rho})~.
\ee
Furthermore, these connections may not be metric compatible, the covariant derivative of the metric tensor is by definition named as
the non-metricity tensor,
\be
Q_{\mu}^{~\rho\sigma}\equiv \hat{\nabla}_{\mu}g^{\rho\sigma}~.
\ee
The torsion and the non-metricity tensors are used to specify the deviations of the geometry of the spacetime from pseudo Riemannian.

Besides the covariant derivative operators, the Riemann tensor is defined the same as that in the metric approach except replacing the Levi-Civita
connections with the general ones,
\be\label{riemann}
\hat{R}^{\sigma}_{~\mu\rho\nu}=\partial_{\nu}\hat{\Gamma}^{\sigma}_{~\rho\mu}-\partial_{\rho}\hat{\Gamma}^{\sigma}_{~\nu\mu}
+\hat{\Gamma}^{\sigma}_{~\nu\kappa}\hat{\Gamma}^{\kappa}_{~\rho\mu}
-\hat{\Gamma}^{\sigma}_{~\rho\kappa}\hat{\Gamma}^{\kappa}_{~\nu\mu}~,
\ee
which has no dependence on the metric. Because $\hat{\Gamma}$ is not symmetric in the sub indices, the Riemann tensor has only one obvious
symmetry: it is antisymmetric in the last two indices.
As the same, the Ricci tensor is defined by the contraction of the first and the third indices in the Riemann tensor
\be\label{ricci}
\hat{R}_{\mu\nu}=\hat{R}^{\sigma}_{~\mu\sigma\nu}~,
\ee
which only depends on the connections and is not necessarily symmetric.
Further contraction of the residual indices through the symmetric metric tensor gives the curvature scalar
\be\label{curvature}
\hat{R}=g^{\mu\nu}\hat{R}_{\mu\nu}~.
\ee

To get the gravitational field equation, we should vary the action with respect to both the metric and the connections.
Naively, the total action is
\be\label{action}
S=S_g+S_m~.
\ee
We only consider the Einstein-Hilbert action for the gravity throughout this paper,
\be\label{einsteinhilbert}
S_g=\frac{M_P^2}{2}\int d^4x\sqrt{g}g^{\mu\nu}\hat{R}_{\mu\nu}(\hat{\Gamma})~.
\ee
The matter fields (represented by $\psi$) may couple to both the metric and the connections directly,
\be
S_m=\int d^4x \sqrt{g} \mathcal{L}_m(\psi,~g_{\mu\nu},~\hat{\Gamma})~.
\ee

Instead of varying the action with respect to $\hat{\Gamma}$ directly, a better way is to do the variation with respect to
the tensor $C^{\mu}_{~\rho\sigma}$ which is expressed as the difference between $\hat{\Gamma}$ and the Levi-Civita connections \cite{Wald},
\be
C^{\mu}_{~\rho\sigma}=\hat{\Gamma}^{\mu}_{~\rho\sigma}-\Gamma^{\mu}_{~\rho\sigma}~.
\ee
As we know, the connections are not tensors but their difference is. The $C$-tensor is also independent of the metric and its variation is equivalent
to that of $\hat{\Gamma}$.
Similarly we can also separate the Ricci tensor $R_{\mu\nu}$ constructed from the Christoffel symbols from the
original Ricci tensor $\hat{R}_{\mu\nu}$,
\be
\hat{R}_{\mu\nu}=R_{\mu\nu}+\nabla_{\nu}C^{\sigma}_{~\sigma\mu}-\nabla_{\sigma}C^{\sigma}_{~\nu\mu}
+C^{\sigma}_{~\nu\rho}C^{\rho}_{~\sigma\mu}-C^{\rho}_{~\rho\sigma}C^{\sigma}_{~\nu\mu}~,
\ee
where the covariant derivative operators without hats are associated with the Levi-Civita connections. With such a separation
the Einstein-Hilbert action we start from becomes
\bea\label{gravity}
S_g&=&\frac{M_P^2}{2}\int d^4x\sqrt{g}g^{\mu\nu}\hat{R}_{\mu\nu}=\frac{M_P^2}{2}\int d^4x\sqrt{g}
[R+g^{\mu\nu}(\nabla_{\nu}C^{\sigma}_{~\sigma\mu}-\nabla_{\sigma}C^{\sigma}_{~\nu\mu}
+C^{\sigma}_{~\nu\rho}C^{\rho}_{~\sigma\mu}-C^{\rho}_{~\rho\sigma}C^{\sigma}_{~\nu\mu})]\nonumber\\
&\rightarrow &\frac{M_P^2}{2}\int d^4x\sqrt{g}
[R+g^{\mu\nu}(C^{\sigma}_{~\nu\rho}C^{\rho}_{~\sigma\mu}-C^{\rho}_{~\rho\sigma}C^{\sigma}_{~\nu\mu})]
~.
\eea
In the third step of the above calculations, we have neglected the terms containing the derivative of the $C$-tensor
because one can find through the Stokes theorem that they only depend on the $C$-tensor and the metric over the boundary
on which all the fields are fixed as implied in the variational principle.
These surface terms have no effects on the dynamics. Please note that the Stokes theorem which works for the covariant derivative without
hat does not work for a general covariant derivative.

If the matter does not couple to the connections directly, the matter Lagrangian is independent of the $C$-tensor. The variation of the action
with respect to the $C$-tensor lead to the following equation,
\be\label{palatini}
-C^{\sigma\alpha}_{~~~\alpha}\delta^{\rho}_{\mu}+C^{\rho\sigma}_{~~~\mu}+C^{\sigma~~\rho}_{~~\mu}-C^{\alpha}_{~~\alpha\mu}g^{\rho\sigma}=0~.
\ee
This is a tensor equation with three independent indices. The contraction of $\rho$ and $\mu$ gives
\be\label{palatini1}
C^{\alpha\sigma}_{~~~\alpha}=C_{\alpha}^{~\alpha\sigma}+3C^{\sigma\alpha}_{~~~\alpha}~,
\ee
and taking the trace on $\rho$ and $\sigma$ we get
\be\label{palatini2}
C^{\alpha}_{~\mu\alpha}=3C^{\alpha}_{~\alpha\mu}+C_{\mu\alpha}^{~~~\alpha}~.
\ee
The contraction of $\sigma$ and $\mu$ in Eq. (\ref{palatini}) leads to the uninteresting identity $0=0$.
From Eqs. (\ref{palatini1}) and (\ref{palatini2}), we have
\bea\label{palatini3}
C^{\alpha~~\sigma}_{~~\alpha}=C^{\sigma\alpha}_{~~~\alpha}=\frac{1}{4}C^{\alpha\sigma}_{~~~\alpha}~,
\eea
with it, Eq. (\ref{palatini}) may be rewritten as
\be\label{eq1}
C^{\rho\sigma\mu}+C^{\sigma\mu\rho}=C^{\sigma\alpha}_{~~~\alpha}g^{\rho\mu}+C^{\mu\alpha}_{~~~\alpha}g^{\rho\sigma}~.
\ee
Permutating the indices $\rho,\sigma,\mu$ give
\bea
& &C^{\mu\rho\sigma}+C^{\rho\sigma\mu}=C^{\rho\alpha}_{~~~\alpha}g^{\mu\sigma}+C^{\sigma\alpha}_{~~~\alpha}g^{\mu\rho}~,\label{eq2}\\
& &C^{\sigma\mu\rho}+C^{\mu\rho\sigma}=C^{\mu\alpha}_{~~~\alpha}g^{\sigma\rho}+C^{\rho\alpha}_{~~~\alpha}g^{\sigma\mu}~.\label{eq3}
\eea
Combining the Eqs. (\ref{eq1}), (\ref{eq2}) and (\ref{eq3}) one obtains finally
\be\label{result}
C^{\sigma\mu\rho}=C^{\mu\alpha}_{~~~\alpha}g^{\rho\sigma}~.
\ee
This result means the $C$-tensor cannot be determined uniquely. So it deserves pointing out that
even in the case that the action of the gravity is Einstein-Hilbert and the matter does not couple to the connections directly,
the metric-affine approach does not have to produce the pseudo-Riemannian spacetime without
further restrictions. This is because the gravity has an extra gauge symmetry, the Einstein-Hilbert action is invariant
under the following projective transformation \cite{Hehl1,Hehl2}
\be\label{projective}
C^{\mu}_{~\rho\sigma}\rightarrow C^{\mu}_{~\rho\sigma}+\delta^{\mu}_{\sigma}\xi_{\rho}~,~~\hat{R}\rightarrow \hat{R}~,
\ee
where $\xi_{\rho}$ is an arbitrary vector field. That means given a solution $C^{\mu}_{~\rho\sigma}$ we can always find other
equivalent solutions $\tilde{C}^{\mu}_{~\rho\sigma}=C^{\mu}_{~\rho\sigma}+\delta^{\mu}_{\sigma}\xi_{\rho}$. Nevertheless this uncertainty does not
bring physical differences because in this case the projective symmetry is only a gauge symmetry. It does not affect the matter field equation and
furthermore from Eqs. (\ref{gravity}), (\ref{palatini3}) and (\ref{result}) that
it has no contribution to the energy-momentum tensor which sources the gravitational field.
If we want to get a unique solution
we should fix the gauge. This can be done by imposing the restrictions on the connections by hand, for example, if we require the spacetime is torsion
free $S^{\mu}_{~\rho\sigma}=0$ or the connections are metric compatible $Q_{\mu}^{~\rho\sigma}=0$, one can show straight forwardly that
the $C$-tensor must vanish and the connections must be the Levi-Civita connections.
Alternatively we can add some additional terms
containing Lagrange multipliers times the connections in the action, the variation of the action with respect to the Lagrange
multipliers will automatically give the constraint equations on the connections. We will discuss this way in more detail in the next section.

\section{Higher derivative scalar field theories with metric-affine variational principle}

The scalar field with covariant higher derivatives must couple to the connections directly. In this paper, for simplicity, we only consider
the scalar field theories in which the Lagrangian density contains the derivatives up to second order.
That is
\be
\mathcal{L}_{\phi}=\mathcal{L}_{\phi}(\phi,~\hat{\nabla}_{\mu}\phi,~\hat{\nabla}_{\mu}\hat{\nabla}_{\nu}\phi)~.
\ee
In the metric approach, all the derivatives are defined through the Levi-Civita connections, so that we can write the Lagrangian density as
\be\label{lc}
\mathcal{L}_{\phi}=\mathcal{L}_{\phi}(\phi,~\phi_{\mu},~\phi_{\nu\mu})~,
\ee
where we have defined
\be
\phi_{\mu}\equiv \nabla_{\mu}\phi~,~~\phi_{\nu\mu}\equiv \nabla_{\mu}\nabla_{\nu}\phi~.
\ee
In the metric-affine formalism, the Lagrangian density depends on the connections $\hat{\Gamma}$
through their associations with the covariant second order derivatives.
With the same strategy used in the previous section, we separate the derivatives associated with the Levi-Civita connections from
those with hats,
\bea
\hat{\nabla}_{\mu}\phi=\phi_{\mu}~,~~
\hat{\nabla}_{\mu}\hat{\nabla}_{\nu}\phi= \phi_{\nu\mu}-
C^{\rho}_{~\mu\nu}\phi_{\rho}~.
\eea
This means the scalar field is equivalently couple to the $C$-tensor directly.
If the $N$th derivative terms with $N>2$ are considered, it is easy to show that the derivatives of the $C$-tensor will be also present in
the Lagrangian density.

Including the gravity, the total action we start from is
\be\label{ef}
S=\frac{M_P^2}{2}\int d^4x\sqrt{g}\hat{R}+
\int d^4 x\sqrt{g}\mathcal{L}_{\phi}(\phi,~\phi_{\mu},~\hat{\nabla}_{\mu}\hat{\nabla}_{\nu}\phi)+S_{ext}~,
\ee
where $S_{ext}$ is the action of other matter which are assumed to be coupled to the gravity only through the metric.
This action is
equivalent to
\be\label{effective}
S=\frac{M_P^2}{2}\int d^4x\sqrt{g}R
+\int d^4x\sqrt{g}[\frac{M_P^2}{2}g^{\mu\nu}(C^{\sigma}_{~\nu\rho}C^{\rho}_{~\sigma\mu}-C^{\rho}_{~\rho\sigma}C^{\sigma}_{~\nu\mu})
+\mathcal{L}_{\phi}(\phi,~\phi_{\mu},~\phi_{\nu\mu}-C^{\rho}_{~\mu\nu}\phi_{\rho})]+S_{ext}~,
\ee
now the second integral at the right hand side of the above equation may be considered as an effective action of the scalar field within the framework
of GR. The whole action does not contain the derivatives of the $C$-tensor, so it represents some constraints
rather than dynamical degrees of freedom. The corresponding constraint equations are obtained by varying the action with respect to the $C$-tensor.
Because in the action the $C$-tensor appears at least quadratically (different from the Lagrange multiplier which also represents a constraint),
its equations actually express the algebraic relations of the $C$-tensor to the
scalar field and its derivatives. Through solving these equations one can find the $C$-tensor expressing as a function of the scalar field $\phi$
and its derivatives.
Then replace $C^{\mu}_{~\rho\sigma}$ in the action (\ref{effective}) with the function found out we may get the the effective Lagrangian
of the scalar field,
\be\label{effective1}
\mathcal{L}'_{\phi}=\frac{M_P^2}{2}g^{\mu\nu}(C^{\sigma}_{~\nu\rho}C^{\rho}_{~\sigma\mu}-C^{\rho}_{~\rho\sigma}C^{\sigma}_{~\nu\mu})+
\mathcal{L}_{\phi}(\phi,~\phi_{\mu},~\phi_{\nu\mu}-C^{\rho}_{~\mu\nu}\phi_{\rho})~,
\ee
where the fact that the $C$-tensor considered as a function of $\phi,~\phi_{\mu}$ and $\phi_{\nu\mu}$ is implied.
Now we can see clearly the difference between the metric approach and the metric-affine approach when the higher derivative terms of the scalar
field are involved. In the metric approach, the scalar field exists in the pseudo-Riemannian spacetime and its dynamics is described by the
Lagrangian density (\ref{lc}). From the metric-affine approach, however, if non-vanishing $C$-tensor is found we can think that the spacetime is
still pseudo-Riemannian but the Lagrangian density of the scalar field gets some modifications or ``corrections".
Of course the non-vanishing $C$-tensor may be also geometrically
interpreted as describing the deviations of the spacetime from pseudo-Riemannian. In this paper we will adopt the first interpretation because it
is more convenient for us to compare the different dynamics produced by these two approaches.

Nevertheless before going ahead, there is a subtlety should be considered for the metric-affine formalism.
The variation of the action (\ref{ef}) or (\ref{effective}) with respect to the $C$-tensor gives the equation
\be\label{cequation}
-C^{\sigma\alpha}_{~~~\alpha}\delta^{\rho}_{\mu}+C^{\rho\sigma}_{~~~\mu}+C^{\sigma~~\rho}_{~~\mu}-
C^{\alpha}_{~~\alpha\mu}g^{\rho\sigma}=\frac{2}{M_P^2}
w^{\rho\sigma}\phi_{\mu}~,
\ee
where
\be
w^{\rho\sigma}\equiv \frac{\partial \mathcal{L}_{\phi}}{\partial (\hat{\nabla}_{\rho}\hat{\nabla}_{\sigma}\phi)}~.
\ee
The left hand side of Eq. (\ref{cequation}) when contracting $\sigma$ and $\mu$ vanishes identically as we have mentioned in the previous
section, this requires
$w^{\rho\mu}\nabla_{\mu}\phi=0$. However this requirement put a very strong constraint on the scalar field and in many cases
it only leads to some trivial solutions which are not interesting for the dark energy or early universe dynamics. In general case
$w^{\rho\mu}\phi_{\mu}$ does not vanish and this makes the whole theory inconsistent. The reason of this inconsistency is that the
Einstein-Hilbert action is invariant under the projective transformation (\ref{projective})
but the general action describing the scalar-connection coupling is not.
To avoid this inconsistency, we should break the projective symmetry in the gravity sector. One choice is to extend the Lagrangian density
of the gravity to include higher order terms like $\hat{R}^{\mu\nu}\hat{R}_{\mu\nu}$ and so forth, but this will introduce extra degrees of freedom.
Another choice is imposing a constraint on the connections by adding in the total action a term $S_L$ which containing
some Lagrange multipliers. This is similar to a gauge fixing in the gauge field theory.
The number of degrees of freedom we need to fix is four, i.e., the
number of the components of the four-vector used for the projective transformation (\ref{projective}).

We will follow Refs. \cite{Hehl2} and \cite{Sotiriou:2006qn} to consider two kinds of Lagrange multipliers which put single vector constraints on the connections.
The first kind of the Lagrange multiplier is
\be
S_{L1}=\int d^4x \sqrt{g}A^{\mu}S_{\mu}~,
\ee
where $S_{\mu}=(1/2)(C^{\rho}_{~\rho\mu}-C^{\rho}_{~\mu\rho})$ is the torsion vector and $A^{\mu}$ is the Lagrange multiplier.
With this additional term Eq. (\ref{cequation}) is modified as
\be\label{cequation1}
-C^{\sigma\alpha}_{~~~\alpha}\delta^{\rho}_{\mu}+C^{\rho\sigma}_{~~~\mu}+C^{\sigma~~\rho}_{~~\mu}-C^{\alpha}_{~\alpha\mu}g^{\rho\sigma}
=\frac{2}{M_P^2}
(w^{\rho\sigma}\phi_{\mu}+\frac{1}{2}\delta^{\sigma}_{\mu}A^{\rho}-\frac{1}{2}\delta^{\rho}_{\mu}A^{\sigma})~.
\ee
The variation with respect to $A^{\mu}$ gives the constraint
\be
S_{\mu}=0~, ~{\rm i.e.,}~C^{\rho}_{~\rho\mu}=C^{\rho}_{~\mu\rho}~.
\ee
These equations are consistent. The contraction of $\sigma$ and $\mu$ in Eq. (\ref{cequation1}) gives
\be
A^{\rho}=-\frac{2}{3}w^{\rho\sigma}\phi_{\sigma}~.
\ee
With the same procedure from Eq. (\ref{palatini}) to Eq. (\ref{result}), we can derive that
\bea\label{final}
C^{\sigma\mu\rho}&=&\frac{1}{M_P^2}
[{1\over 2}(-w^{\sigma\alpha}\phi_{\alpha}-w^{\alpha\sigma}\phi_{\alpha}+w^{\alpha}_{~\alpha}\phi^{\sigma})g^{\rho\mu}
+{1\over 6}(3w^{\mu\alpha}\phi_{\alpha}-w^{\alpha\mu}\phi_{\alpha}-3w^{\alpha}_{~\alpha}\phi^{\mu})g^{\rho\sigma}\nonumber\\
&+&{1\over 6}(-w^{\rho\alpha}\phi_{\alpha}+3w^{\alpha\rho}\phi_{\alpha}-3w^{\alpha}_{~\alpha}\phi^{\rho})g^{\sigma\mu}
+\phi^{\mu} w^{\rho\sigma}+\phi^{\rho} w^{\sigma\mu}-\phi^{\sigma} w^{\mu\rho}]~.
\eea
From the above equation we can solve for the $C$-tensor as a function of $\phi$ and its derivatives, then
substitute this function in Eq. (\ref{effective1}), we will get the desired effective Lagrangian for the higher derivative scalar field.
We see from Eq. (\ref{final}) that the $C$-tensor is suppressed at least by the square of the Planck mass, so in the resulting effective
Lagrangian density (\ref{effective1}) the modification terms are also suppressed at least by $M_P^2$.

The second kind of the Lagrange multiplier we consider in this paper is described by
\be
S_{L2}=\int d^4x \sqrt{g}B^{\mu}Q_{\mu}~,
\ee
where $Q_{\mu}=(1/4)Q_{\mu\rho}^{~~~\rho}=(1/2)C^{\rho}_{~\mu\rho}$ is the Weyl vector and $B^{\mu}$ is the Lagrange multiplier.
With this, the equation for the $C$-tensor becomes
\be\label{cequation2}
-C^{\sigma\alpha}_{~~~\alpha}\delta^{\rho}_{\mu}+C^{\rho\sigma}_{~~~\mu}+C^{\sigma~~\rho}_{~~\mu}-C^{\alpha}_{~\alpha\mu}g^{\rho\sigma}
=\frac{2}{M_P^2}
(w^{\rho\sigma}\phi_{\mu}-\frac{1}{2}\delta^{\sigma}_{\mu}B^{\rho})~,
\ee
and the constraint put by the Lagrange multiplier is
\be
Q_{\mu}={1\over 2}C^{\rho}_{~\mu\rho}=0~.
\ee
Similarly we get the result
\bea\label{final2}
C^{\sigma\mu\rho}&=&\frac{1}{M_P^2}
[{1\over 4}(-w^{\sigma\alpha}\phi_{\alpha}-2w^{\alpha\sigma}\phi_{\alpha}+2w^{\alpha}_{~\alpha}\phi^{\sigma})g^{\rho\mu}
+{1\over 8}(3w^{\mu\alpha}\phi_{\alpha}-2w^{\alpha\mu}\phi_{\alpha}-2w^{\alpha}_{~\alpha}\phi^{\mu})g^{\rho\sigma}\nonumber\\
&+&{1\over 4}(-w^{\rho\alpha}\phi_{\alpha}+2w^{\alpha\rho}\phi_{\alpha}-2w^{\alpha}_{~\alpha}\phi^{\rho})g^{\sigma\mu}
+\phi^{\mu} w^{\rho\sigma}+\phi^{\rho} w^{\sigma\mu}-\phi^{\sigma} w^{\mu\rho}]~.
\eea
Comparing with Eq. (\ref{final}), this result only have the differences in some numerical factors.

\section{A simple model}

Now we use a simple model to illustrate whether the differences caused by the metric-affine formalism are significant. The model
has a Lagrangian density degenerate in the higher derivative terms,
\be
\mathcal{L}_{\phi}=f(\phi,~X)\widehat{\Box}\phi~,
\ee
where $X=\frac{1}{2}\phi_{\mu}\phi^{\mu}$, $f$ does not depend on the second derivative of the scalar field and the d'Alembert operator is defined as
$\widehat{\Box}\equiv g^{\mu\nu}\hat{\nabla}_{\mu}\hat{\nabla}_{\nu}$. For the general connections we should be careful about the order
of the metric and the derivative operators.
We can find that $w^{\rho\sigma}$ is symmetric
\be
w^{\rho\sigma}=fg^{\rho\sigma}~.
\ee
With the first kind of Lagrange multiplier, the result (\ref{final}) is simply
\be
C^{\sigma\mu\rho}=-\frac{2f}{3M_P^2}(\phi^{\rho}g^{\sigma\mu}+\phi^{\mu}g^{\rho\sigma})~.
\ee
In geometric language, because $C^{\sigma\mu\rho}$ is symmetric about $\mu$ and $\rho$, the torsion tensor vanishes.
But the non-metricity tensor is non-zero,
\be
Q_{\mu}^{~\rho\sigma}=-\frac{2f}{3M_P^2}(\phi^{\rho}\delta^{\sigma\mu}+\phi^{\sigma}\delta^{\rho\mu}+2\phi_{\mu}g^{\rho\sigma})~.
\ee
The effective Lagrangian density what we need is
\be
\mathcal{L}_{\phi}'=f\Box\phi+\frac{4f^2}{3M_P^2}X~.
\ee
So, we can see from the above equation that comparing with the metric approach the metric-affine formalism brings the extra term
$\frac{4f^2}{3M_P^2}X$ to the Lagrangian density. This modification term is suppressed by the square of the Planck mass and in many cases cause minor
differences in the dynamics. However, if the original Lagrangian itself relies on the higher dimensional operators, this modification may have important
effects. For example, consider the model in which
\be\label{this}
f=\frac{\phi}{2}+\frac{c}{M_P^3} X~,
\ee
where $c$ is a dimensionless constant.
This model has been studied with the metric approach in Refs. \cite{Deffayet:2010qz,Li:2011qf} for addressing the issue of dark energy and
in Ref. \cite{Qiu:2011cy} for bouncing universe.
With the metric approach this model is symmetric under the field shift $\phi\rightarrow \phi+C$ with constant
$C$. Correspondingly the conserved current is
\be
J^{\mu}=\frac{c}{M_P^3}(\nabla^{\mu}X-\phi^{\mu}\Box\phi)+\phi^{\mu}~.
\ee
In the spatially-flat Friedmann-Robertson-Walker universe, the conservation law demands that the
``charge density" $J^0=\dot\phi-\frac{3c}{M_P^3}H\dot\phi^2$
scales as $a^{-3}$, where $a$ is the scale factor of the universe, $H$ is the Hubble parameter and the dot represents the derivative with respect to
time. This makes the analysis of the dynamics gets
large simplification. Here we only review the main steps, for more details see Ref. \cite{Li:2011qf}.
In the expanding universe $J^0$ dilutes quickly and the equation of state of the scalar field is asymptotically
$w\simeq \frac{2\dot H}{3H^2}-1=-(2+w_b)$, where $w_b$ is the equation of state of the component dominating the universe.
So in the matter dominated era $w=-2$ and if the universe is dominated by the scalar field itself $w=-1$.
As the dark energy this model can account for how the universe shifts from the matter domination to the phase of the accelerating
expansion. This dark energy model has a feature that the universe will end up in the de Sitter space.

However if we start from the same Lagrangian density but adopt the metric-affine formalism, the model (\ref{this}) is not shift symmetric.
We can see that the effective Lagrangian density is
\be\label{that}
\mathcal{L}_{\phi}'=\mathcal{L}_{\phi}+\frac{1}{3M_P^2}X\phi^2+\frac{4c}{3M_P^5}X^2(\phi+\frac{c}{M_P^3}X)=
(\frac{\phi}{2}+\frac{c}{M_P^3} X)\Box\phi+\frac{1}{3M_P^2}X\phi^2+\frac{4c}{3M_P^5}X^2(\phi+\frac{c}{M_P^3}X)~.
\ee
Among the three modification terms, the last two terms are suppressed by much higher order powers of the Planck mass and are expected to
have negligible effects. But the first correction term $\frac{1}{3M_P^2}X\phi^2$ has lower order than the higher dimensional term
$\frac{c}{M_P^3} X\Box\phi$ originally exists in
the Lagrangian we start from. In fact this term can be considered as an effective mass for the scalar field. Its effect on dark energy has
been investigated in Ref. \cite{Li:2011qf}.  With this effective mass the whole picture of the future universe is changed.
The universe will not end up in the de Sitter space, the equation of state of the
scalar field will increase from $-1$ to $1$, then its energy density will decrease very fast, faster than the energy density of the dark matter, and the
universe will return to the phase of matter domination. It deserves pointing out that such like effective mass
has also been introduced in inflation model \cite{Takahashi:2010ky,Nakayama:2010kt} to suppress the amplitudes of the primordial perturbations.


\section{Summary}

In this paper we studied the higher derivative scalar fields which received much attention in addressing the issues of the early and late universe
with the metric-affine formalism.
As a covariant theory, the higher derivative scalar fields have to couple to the connections directly and as a consequence
within the metric-affine formalism they have dynamics different from those within the metric formalism.
We showed by the method of effective Lagrangian that usually these differences are suppressed by the
powers of the Planck mass and are expected to be small. However, if the Lagrangian density we start from itself relies on the higher dimensional
operators the modifications brought by the metric-affine formalism might introduce important deviations.
We used a degenerate higher derivative dark energy model to illustrate that these two approaches lead to significantly different pictures of the
future universe. In other applications of the higher derivative scalar fields, for examples the inflation models, the non-singular bouncing universe,
and the covariantization of the Galileons, we believe the metric-affine formalism as an alternative to the metric formalism deserves more attention.

\section{Acknowledgement}
This work is supported in part by National Science
Foundation of China under Grants No. 11075074 and No. 11065004, by
the Specialized Research Fund for the Doctoral Program of Higher
Education (SRFDP) under Grant No. 20090091120054 and by SRF for
ROCS, SEM.

{}


\begin{thebibliography}{}

\bibitem{Guth:1980zm}
  A.~H.~Guth,
  Phys.\ Rev.\ D {\bf 23} (1981) 347.

\bibitem{Linde:1981mu}
  A.~D.~Linde,
  Phys.\ Lett.\ B {\bf 108} (1982) 389.

\bibitem{Albrecht:1982wi}
  A.~Albrecht and P.~J.~Steinhardt,
  Phys.\ Rev.\ Lett.\  {\bf 48} (1982) 1220.  

\bibitem{Novello:2008ra}
  M.~Novello and S.~E.~P.~Bergliaffa,
  hys.\ Rept.\  {\bf 463} (2008) 127  [arXiv:0802.1634 [astro-ph]].  

\bibitem{Steinhardt:2002ih}
  P.~J.~Steinhardt and N.~Turok,
  Science {\bf 296} (2002) 1436.  

\bibitem{Gasperini:1992em}
  M.~Gasperini and G.~Veneziano,
  Astropart.\ Phys.\  {\bf 1} (1993) 317  [hep-th/9211021].  

\bibitem{Copeland:2006wr}
  E.~J.~Copeland, M.~Sami and S.~Tsujikawa,
  Int.\ J.\ Mod.\ Phys.\ D {\bf 15} (2006) 1753  [hep-th/0603057].  

\bibitem{Li:2005fm}
  M.~-z.~Li, B.~Feng and X.~-m.~Zhang,
  JCAP {\bf 0512} (2005) 002  [hep-ph/0503268].  

\bibitem{Feng:2004ad}
  B.~Feng, X.~-L.~Wang and X.~-M.~Zhang,
  Phys.\ Lett.\ B {\bf 607} (2005) 35  [astro-ph/0404224].  

\bibitem{Anisimov:2005ne}
  A.~Anisimov, E.~Babichev and A.~Vikman,
  JCAP {\bf 0506} (2005) 006  [astro-ph/0504560].  

\bibitem{Cai:2007qw}
  Y.~-F.~Cai, T.~Qiu, Y.~-S.~Piao, M.~Li and X.~Zhang,
  JHEP {\bf 0710} (2007) 071  [arXiv:0704.1090 [gr-qc]].  

\bibitem{Cai:2008qw}
  Y.~-F.~Cai, T.~-t.~Qiu, R.~Brandenberger and X.~-m.~Zhang,
  Phys.\ Rev.\ D {\bf 80} (2009) 023511  [arXiv:0810.4677 [hep-th]].  

\bibitem{Nicolis:2008in}
  A.~Nicolis, R.~Rattazzi and E.~Trincherini,
  Phys.\ Rev.\  D {\bf 79} (2009) 064036
  [arXiv:0811.2197 [hep-th]].

\bibitem{Dvali:2000hr}
  G.~R.~Dvali, G.~Gabadadze and M.~Porrati,
  Phys.\ Lett.\  B {\bf 485} (2000) 208
  [arXiv:hep-th/0005016].

\bibitem{Deffayet:2009wt}
  C.~Deffayet, G.~Esposito-Farese and A.~Vikman,
  Phys.\ Rev.\ D {\bf 79} (2009) 084003  [arXiv:0901.1314 [hep-th]].  


\bibitem{Horndeski:1974}
 G.~W.~Horndeski,
 Int.\ J.\ Theor.\ Phys. {\bf 10} (1974) 363.

\bibitem{Li:2011qf}
  M.~Li, T.~Qiu, Y.~Cai and X.~Zhang,
  JCAP {\bf 1204} (2012) 003
  [arXiv:1112.4255 [hep-th]].

\bibitem{Deffayet:2010qz}
  C.~Deffayet, O.~Pujolas, I.~Sawicki and A.~Vikman,
  JCAP {\bf 1010} (2010) 026
  [arXiv:1008.0048 [hep-th]].

\bibitem{Pujolas:2011he}
  O.~Pujolas, I.~Sawicki and A.~Vikman,
  JHEP {\bf 1111} (2011) 156
  [arXiv:1103.5360 [hep-th]].

\bibitem{Kobayashi:2010cm}
  T.~Kobayashi, M.~Yamaguchi, J.~Yokoyama,
  Phys.\ Rev.\ Lett.\  {\bf 105} (2010) 231302
  [arXiv:1008.0603 [hep-th]]; \\
%
  K.~Kamada, T.~Kobayashi, M.~Yamaguchi, J.~Yokoyama,
  Phys.\ Rev.\  D {\bf 83} (2011) 083515
  [arXiv:1012.4238 [astro-ph.CO]].

\bibitem{Qiu:2011cy}
  T.~Qiu, J.~Evslin, Y.~F.~Cai, M.~Li and X.~Zhang,
  JCAP {\bf 1110} (2011) 036
  [arXiv:1108.0593 [hep-th]];\\
%
  D.~A.~Easson, I.~Sawicki and A.~Vikman,
  JCAP {\bf 1111} (2011) 021  [arXiv:1109.1047 [hep-th]].  


\bibitem{deRham:2010eu}
  C.~de Rham and A.~J.~Tolley,
  JCAP {\bf 1005} (2010) 015
  [arXiv:1003.5917 [hep-th]]; \\
%
  C.~Burrage, C.~de Rham, D.~Seery and A.~J.~Tolley,
  JCAP {\bf 1101} (2011) 014
  [arXiv:1009.2497 [hep-th]].

%
\bibitem{Deffayet:2009mn}
  C.~Deffayet, S.~Deser and G.~Esposito-Farese,
  Phys.\ Rev.\  D {\bf 80} (2009) 064015
  [arXiv:0906.1967 [gr-qc]];\\
%
  C.~Deffayet, X.~Gao, D.~A.~Steer and G.~Zahariade,
  Phys.\ Rev.\  D {\bf 84} (2011) 064039
  [arXiv:1103.3260 [hep-th]].


\bibitem{Gao:2011qe}
  X.~Gao and D.~A.~Steer,
  JCAP {\bf 1112} (2011) 019  [arXiv:1107.2642 [astro-ph.CO]].

\bibitem{Chow:2009fm}
  N.~Chow and J.~Khoury,
  Phys.\ Rev.\  D {\bf 80} (2009) 024037
  [arXiv:0905.1325 [hep-th]]; \\
%
  J.~Khoury, J.~L.~Lehners and B.~A.~Ovrut,
  Phys.\ Rev.\  D {\bf 84} (2011) 043521
  [arXiv:1103.0003 [hep-th]].

\bibitem{Silva:2009km}
  F.~P.~Silva and K.~Koyama,
  Phys.\ Rev.\  D {\bf 80} (2009) 121301
  [arXiv:0909.4538 [astro-ph.CO]]; \\
%
  S.~Mizuno and K.~Koyama,
  Phys.\ Rev.\  D {\bf 82} (2010) 103518
  [arXiv:1009.0677 [hep-th]].

\bibitem{DeFelice:2010pv}
  A.~De Felice and S.~Tsujikawa,
  Phys.\ Rev.\ Lett.\  {\bf 105} (2010) 111301
  [arXiv:1007.2700 [astro-ph.CO]]; \\
%
  S.~Nesseris, A.~De Felice and S.~Tsujikawa,
  Phys.\ Rev.\  D {\bf 82} (2010) 124054
  [arXiv:1010.0407 [astro-ph.CO]]; \\
%
  A.~De Felice, R.~Kase and S.~Tsujikawa,
  Phys.\ Rev.\  D {\bf 83} (2011) 043515
  [arXiv:1011.6132 [astro-ph.CO]].


\bibitem{Hinterbichler:2010xn}
  K.~Hinterbichler, M.~Trodden and D.~Wesley,
  Phys.\ Rev.\  D {\bf 82} (2010) 124018
  [arXiv:1008.1305 [hep-th]]; \\
%
  G.~L.~Goon, K.~Hinterbichler and M.~Trodden,
  Phys.\ Rev.\  D {\bf 83} (2011) 085015
  [arXiv:1008.4580 [hep-th]].

\bibitem{Creminelli:2010ba}
  P.~Creminelli, A.~Nicolis and E.~Trincherini,
  JCAP {\bf 1011} (2010) 021
  [arXiv:1007.0027 [hep-th]]; \\
%
  A.~Padilla, P.~M.~Saffin and S.~Y.~Zhou,
  Phys.\ Rev.\  D {\bf 83} (2011) 045009
  [arXiv:1008.0745 [hep-th]]; \\
%
  P.~Creminelli, G.~D'Amico, M.~Musso, J.~Norena and E.~Trincherini,
  JCAP {\bf 1102} (2011) 006
  [arXiv:1011.3004 [hep-th]]; \\
%
  M.~Wyman,
  Phys.\ Rev.\ Lett.\  {\bf 106} (2011) 201102
  [arXiv:1101.1295 [astro-ph.CO]]; \\
%
  K.~Van Acoleyen and J.~Van Doorsselaere,
  Phys.\ Rev.\  D {\bf 83} (2011) 084025
  [arXiv:1102.0487 [gr-qc]]; \\
%
  L.~P.~Levasseur, R.~Brandenberger and A.~C.~Davis,
  Phys.\ Rev.\  D {\bf 84} (2011) 103512
  arXiv:1105.5649 [astro-ph.CO]; \\
%
  S.~Renaux-Petel,
  Class.\ Quant.\ Grav.\  {\bf 28} (2011) 182001
  [Erratum-ibid.\  {\bf 28}, (2011) 249601]
  arXiv:1105.6366 [astro-ph.CO]; \\
%
  Z.~G.~Liu, J.~Zhang and Y.~S.~Piao,
  Phys.\ Rev.\  D {\bf 84} (2011) 063508
  arXiv:1105.5713 [astro-ph.CO]; \\
%
  X.~Gao,
  JCAP {\bf 1110} (2011) 021
  arXiv:1106.0292 [astro-ph.CO]; \\
%
  J.~Evslin, T.~Qiu,
  JHEP {\bf 1111} (2011) 032
  [arXiv:1106.0570 [hep-th]]; \\
%
  S.~Renaux-Petel,
  arXiv:1107.5020 [astro-ph.CO];\\
%
 H.~Wang, T.~Qiu and Y.~-S.~Piao,
  Phys.\ Lett.\ B {\bf 707} (2012) 11  [arXiv:1110.1795 [hep-ph]].

\bibitem{Hehl1}
F.~W.~Hehl and G.~D.~Kerling, Gen.\ Rel.\ Grav.\ {9} (1978) 691.

\bibitem{Hehl2}
F.~W.~Hehl, E.~A.~Lord and L.~L.~Smalley, Gen.\ Rel.\ Grav.\ {13} (1981) 1037.

\bibitem{Vollick:2003aw}
  D.~N.~Vollick,
  Phys.\ Rev.\ D {\bf 68} (2003) 063510  [astro-ph/0306630].  

\bibitem{Sotiriou:2006qn}
  T.~P.~Sotiriou and S.~Liberati,
  Annals Phys.\  {\bf 322} (2007) 935  [gr-qc/0604006].  

\bibitem{Bauer:2008zj}
  F.~Bauer and D.~A.~Demir,
  Phys.\ Lett.\ B {\bf 665} (2008) 222  [arXiv:0803.2664 [hep-ph]].  

\bibitem{weyl}
H.~Weyl, Phys.\ Rev.\ {\bf 77} (1950) 699;\\
S.~Deser and B.~Zumino, Phys.\ Lett.\ B {\bf 62} (1976) 335.


\bibitem{Wald}
R.~M.~Wald, General Relativity, The University of Chicago Press, Chicago, 1984.

\bibitem{Takahashi:2010ky}
  F.~Takahashi,
  Phys.\ Lett.\ B {\bf 693} (2010) 140  [arXiv:1006.2801 [hep-ph]].  

\bibitem{Nakayama:2010kt}
  K.~Nakayama and F.~Takahashi,
  JCAP {\bf 1011} (2010) 009  [arXiv:1008.2956 [hep-ph]].  

\end{thebibliography}
\end{document}